\documentclass[aps,12pt,showpacs,preprint,groupedaddress,floatfix]{revtex4}
\usepackage{graphicx}
\usepackage{dcolumn}
\usepackage{bm}
\usepackage{amssymb}

\newcommand{\tfrac}[2]{{\textstyle {#1\over #2}}}

\newcommand{\bra}[1]{\left\langle #1 \right|}
\newcommand{\ket}[1]{\left| #1 \right\rangle}

\begin{document}

\title{Beyond the relativistic mean-field approximation: 
configuration mixing of angular momentum projected 
wave functions}
\author{T. Nik\v si\' c}
\author{D. Vretenar}
\affiliation{Physics Department, Faculty of Science, University of Zagreb, 
Croatia, and \\
Physik-Department der Technischen Universit\"at M\"unchen, D-85748 Garching,
Germany}
\author{P. Ring}
\affiliation{Physik-Department der Technischen Universit\"at M\"unchen, 
D-85748 Garching,
Germany}
\date{\today}

\begin{abstract}
We report the first study of restoration of rotational symmetry 
and fluctuations of the quadrupole deformation in the framework 
of relativistic mean-field models. A model is developed which 
uses the generator coordinate method to perform
configuration mixing calculations of angular momentum projected
wave functions, calculated in a relativistic point-coupling model.
The geometry is restricted to axially symmetric shapes, and
the intrinsic wave functions are generated from the solutions of 
the constrained relativistic mean-field + BCS equations in an 
axially deformed oscillator basis. A number of illustrative 
calculations are performed for the nuclei $^{194}$Hg and $^{32}$Mg,
in comparison with results obtained in non-relativistic models 
based on Skyrme and Gogny effective interactions.
\end{abstract}

\maketitle

\section{\label{secI}Introduction}
The rich variety of nuclear shapes far from stability has been 
the subject of extensive experimental and theoretical studies. 
The variation of ground-state shapes in an isotopic chain, for
instance, is governed by the evolution of shell structure. 
In particular, far from the $\beta$-stability line the 
energy spacings between single-particle levels change considerably 
with the number of neutrons and/or protons. This can result in 
reduced spherical shell gaps, modifications of shell structure, 
and in some cases spherical magic numbers may disappear. For example, 
in neutron-rich nuclei $N=6,16,34...$ can become magic numbers, whereas
$N=8,20,28...$ disappear. The reduction of a spherical shell closure 
is associated with the occurrence of deformed ground states and, in 
a number of cases, with the phenomenon of shape coexistence. 

Both the global shell-model approach and self-consistent 
mean-field models have been employed in the description of shell 
evolution far from stability. The basic advantages of the shell 
model is the ability to describe simultaneously all spectroscopic
properties of low-lying states for a large domain of nuclei, 
effective interactions that can be related to two- 
and three-nucleon bare forces, and a description of collective
properties in the laboratory frame. On the other hand, since 
effective interactions strongly depend on the choice of active 
shells and truncation schemes, there is no universal shell-model
interaction that can be used for all nuclei. Moreover, because a
large number of two-body matrix elements has to be adjusted to data, 
extrapolations to exotic systems far from stability cannot be very
reliable. Heavy exotic nuclei with very large valence spaces require
calculations with matrix dimensions that are far beyond the limits 
of current shell model variants. 

Properties of heavy nuclei with a large number of active valence 
nucleons are best described in the framework of self-consistent 
mean-field models. A variety of structure phenomena, not only in 
medium-heavy and heavy stable nuclei, but also in regions of exotic
nuclei far from the line of $\beta$-stability and close to the nucleon
drip-lines, have been successfully described with mean-field 
models based on the Gogny interaction, the Skyrme energy functional,
and the relativistic meson-exchange effective Lagrangian 
\cite{BHR.03,VALR.05}. The self-consistent mean-field approach to 
nuclear structure represents an approximate implementation of 
Kohn-Sham density functional theory, which enables a description of
the nuclear many-body problem in terms of a universal energy density
functional. This framework, extended to take into account 
the most important correlations, provides a detailed microscopic 
description of structure phenomena associated with shell evolution
in exotic nuclei. When compared to the shell model, important advantages
of the mean-field approach include the use of global effective nuclear 
interactions, the treatment of arbitrarily heavy systems including 
superheavy elements, and the intuitive picture of intrinsic shapes. 

A quantitative description of shell evolution, in particular 
the treatment of shape coexistence phenomena, necessitates the 
inclusion of many-body correlations beyond the mean-field
approximation. The starting point is usually a constrained Hartree-Fock 
plus BCS (HFBCS), or Hartree-Fock-Bogoliubov (HFB) calculation of the 
potential energy surface with the mass quadrupole components as 
constrained quantities. In most applications calculations have 
been restricted to axially symmetric, parity conserving configurations.
The erosion of spherical shell-closures in nuclei far from stability 
leads to deformed intrinsic states and, in some cases, mean-field 
potential energy surfaces with almost degenerate prolate and oblate 
minima. To describe nuclei with soft potential energy surfaces and/or
small energy differences between coexisting minima, it is necessary
to explicitly consider correlation effects beyond the mean-field level. 
The rotational energy correction, i.e. the energy gained by the restoration
of rotational symmetry, is proportional to the quadrupole deformation of 
the intrinsic state and can reach several MeV for a well deformed configuration.
Fluctuations of quadrupole deformation also contribute to the 
correlation energy. Both types of correlations can be included 
simultaneously by mixing angular momentum projected states corresponding 
to different quadrupole moments. The most effective approach for
configuration mixing calculations is the generator coordinate method (GCM),
with multipole moments used as coordinates that generate the intrinsic wave
functions.

In a series of recent papers \cite{RER.02a,GER.02,RER.03,RER.04}, 
the angular momentum projected GCM with the axial quadrupole moment as 
the generating coordinate, and intrinsic configurations calculated  
in the HFB model with the finite range Gogny interaction, has been applied
in studies of shape-coexistence phenomena that result from the erosion 
of the $N=20$ and $N=28$ spherical shells in neutron-rich nuclei. 
Good agreement with experimental data has been obtained for the $2^+$ 
excitation energies, and B(E2) transition probabilities of the 
$N=28$ neutron-rich isotones \cite{RER.02a}. The systematic study of 
the ground and low-lying excited states of the 
even-even $^{20-40}$Mg \cite{GER.02} is particularly interesting, because
this chain of isotopes includes three spherical magic numbers 
$N=8, 20, 28$. It has been shown that the $N=8$ shell closure is 
preserved, whereas deformed ground states are calculated for $N=20$ and 
$N=28$. In particular, the ground state of $^{32}$Mg becomes deformed 
as a result of a fine balance between the energy 
correction associated with the restoration of rotational symmetry and
the correlations induced by quadrupole fluctuations. In a similar  
analysis of the chain of even-even isotopes $^{20-34}$Ne \cite{RER.03},
it has been shown that the ground state of the $N=20$ nucleus $^{30}$Ne 
is deformed, but less than the ground state of its isotone $^{32}$Mg. 
The model has recently been applied in an analysis of shape 
coexistence and quadrupole collectivity in the neutron-deficient 
Pb isotopes \cite{RER.04}. 
A good qualitative agreement with available data has been
found, especially for rotational bands built on coexisting 
low-lying oblate and prolate states.

Another very sophisticated model \cite{VHB.00} 
which extends the self-consistent mean-field approach
by including correlations, is based on constrained HF+BCS 
calculations with Skyrme effective interactions in the particle-hole 
channel and a density-dependent contact force in the pairing channel.
Particle number and rotational symmetry are restored by projecting 
self-consistent mean-field wave functions on the correct numbers 
of neutrons and protons, and on angular momentum. Finally, a mixing 
of the projected wave functions corresponding to different quadrupole 
moments is performed with a discretized version of the generator 
coordinate method. The model has been successfully tested in 
the study of shape coexistence in $^{16}$O \cite{BH.03}, and in the
analysis of the coexistence of spherical, deformed, and superdeformed 
states in $^{32}$S, $^{36}$Ar, $^{38}$Ar and $^{40}$Ca \cite{BFH.03}.
For the doubly-magic nucleus $^{16}$O 
this parameter-free approach provides a very good description of 
those low-spin states which correspond to axially and 
reflection-symmetric shapes, and allows the interpretation of 
their structure in terms of self-consistent $np-nh$ states. 
A very important recent application is the study of low-lying 
collective excitation spectra of the neutron-deficient lead isotopes
$^{182-194}$Pb \cite{DBBH.03,BBDH.04}. A configuration mixing 
of angular-momentum and particle-number projected self-consistent 
mean-field states, calculated with the Skyrme SLy6 effective interaction, 
qualitatively reproduces the coexistence of spherical, oblate, prolate, 
and superdeformed prolate structures in neutron-deficient Pb nuclei.  

Even though the self-consistent relativistic mean-field (RMF) framework 
has been employed in many studies of deformed nuclei,
applications of meson-exchange and point-coupling models have so far
been restricted to the mean-field level. In this work we report the
first study of restoration of rotational symmetry and fluctuations
of the quadrupole deformation in the framework of relativistic 
mean-field models. We perform a GCM configuration mixing of angular 
momentum projected wave functions that are calculated in a relativistic 
point-coupling model.

In Section \ref{secII} we present an outline of the relativistic 
point-coupling model which will be used to generate mean-field 
wave functions with axial symmetry, introduce the formalism of
the generator coordinate method, and describe in detail the 
procedure of configuration mixing of angular momentum projected
wave functions. In Section \ref{secIII} our model for GCM 
configuration mixing is investigated in a study of quadrupole dynamics 
in the nucleus $^{194}$Hg, and $^{32}$Mg is used as a test case 
for the configuration mixing calculation of angular momentum 
projected states. Section \ref{secIV} 
summarizes the results of the present investigation and 
ends with an outlook for future studies.

\section{\label{secII}Configuration mixing of angular momentum projected
mean-field wave functions}
In this section we review the self-consistent relativistic point-coupling 
model which will be used to generate constrained mean-field states, and  
the solution of the corresponding single-nucleon Dirac equation in an
axially symmetric harmonic oscillator basis. Starting with a short outline
of the generator coordinate method, we describe the technical details 
of the configuration mixing of angular momentum projected wave functions.  

\subsection{\label{subIIa}The relativistic point-coupling model}
Most applications of the self-consistent relativistic mean-field 
framework have used the finite-range meson-exchange 
representation, in which the nucleus is described as a system of Dirac 
nucleons coupled to exchange mesons and the electromagnetic field 
through an effective Lagrangian. A medium dependence of the 
effective nuclear interaction can be introduced 
either by including non-linear meson self-interaction terms
in the Lagrangian, or by assuming an explicit density dependence 
for the meson-nucleon couplings \cite{VALR.05}.
An alternative representation is formulated in terms 
of point-coupling (PC) (contact) nucleon-nucleon interactions
\cite{MNH.92,Hoch.94,FML.96,RF.97,BMM.02}. In RMF-PC models
the medium dependence of the interaction can be taken into 
account by the inclusion of higher order interaction terms, 
for instance six-nucleon vertices 
$(\bar\psi\psi)^3$, and eight-nucleon vertices $(\bar\psi\psi)^4$ and
$[(\bar\psi\gamma_\mu\psi)(\bar\psi\gamma^\mu\psi)]^2$, or it 
can be encoded in the effective couplings, i.e. in the strength 
parameters of the interaction in the isoscalar and isovector channels. 
When employed in studies of ground-state properties of finite nuclei, 
the two representations produce results of comparable quality. 
The point-coupling formulation, however, avoids some of the 
constraints imposed in the meson-exchange picture as, for instance, 
the use of the fictitious sigma-meson in the isoscalar-scalar channel.
The self-consistent PC models are also closer in spirit to the 
nuclear density functional theory, in which the exact energy functional, 
including higher-order correlations, is approximated with powers 
and gradients of ground-state nucleon densities. The point-coupling 
representation, with medium-dependent vertex functions, provides a 
natural framework in which chiral effective field theory can be 
employed to construct the nuclear energy density functional, thus
establishing a link between the rich nuclear phenomenology and 
the underlying microscopic theory of low-energy QCD \cite{FKV.03,FKV.04}.

A detailed description of the point-coupling model that we use 
in this work can be found, for instance, in Ref. \cite{BMM.02}, 
together with a thorough discussion of the choice of various 
parameter sets that determine the effective interactions. 
Here we only outline the essential features of the model and 
of its mean-field solution for a deformed axially symmetric nucleus.

The relativistic point-coupling Lagrangian is
built from basic densities and currents bilinear in the Dirac
spinor field $\psi$ of the nucleon:
\begin{equation}
  {\bar{\psi}} {\mathcal O}_\tau \Gamma  {\psi}
  \quad,\quad
  {\mathcal O}_\tau\in\{ {1},\tau_i\}
  \quad,\quad
  \Gamma\in\{1,\gamma_\mu,\gamma_5,\gamma_5\gamma_\mu,\sigma_{\mu\nu}\}\; .
\end{equation}
Here $\tau_i$ are the isospin Pauli matrices and
$\Gamma$ generically denotes the Dirac matrices.
The interaction terms of the Lagrangian are products of these
bilinears. Although a general effective Lagrangian
can be written as a power series in the currents 
${\bar{\psi}} {\mathcal O}_\tau \Gamma  {\psi}$ and their derivatives,
it is well known from numerous applications
of relativistic mean-field models that properties of symmetric and asymmetric
nuclear matter, as well as empirical ground state properties of finite
nuclei, constrain only the isoscalar-scalar (S), the isoscalar-vector (V),
the isovector-vector (TV), and to a certain extent the isovector-scalar (TS)
channels. In this work we consider a model with four-, six-, and eight-fermion 
point couplings (contact interactions) \cite{BMM.02}, defined 
by the Lagrangian density:
\begin{equation}
\begin{array}{lcl}
  {\cal L} 
  & = & 
  {\cal L}^{\rm free} + {\cal L}^{\rm 4f} + {\cal L}^{\rm hot}
  + {\cal L}^{\rm der} + {\cal L}^{\rm em},
\\[12pt]
  {\cal L}^{\rm free} \hfill
  & = & 
  \bar\psi ({\rm i}\gamma_\mu\partial^\mu -m)\psi,
\\[6pt]
  {\cal L}^{\rm 4f} \hfill
  & = & 
  - \tfrac{1}{2}\, \alpha_{\rm S} (\bar\psi\psi)(\bar\psi\psi)
  - \tfrac{1}{2}\, 
    \alpha_{\rm V}(\bar\psi\gamma_\mu\psi)(\bar\psi\gamma^\mu\psi)
\\[6pt]
  & &
   - \tfrac{1}{2}\, \alpha_{\rm TS} (\bar\psi\vec\tau\psi) \cdot
   (\bar\psi\vec\tau\psi)
  - \tfrac{1}{2}\,  \alpha_{\rm TV} (\bar\psi\vec\tau\gamma_\mu\psi)
    \cdot (\bar\psi\vec\tau\gamma^\mu\psi),
\\[6pt]
  {\cal L}^{\rm hot} 
  & = &  
  - \tfrac{1}{3}\, \beta_{\rm S} (\bar\psi\psi)^3 - \tfrac{1}{4}\, 
    \gamma_{\rm S} (\bar\psi\psi)^4 - \tfrac{1}{4}\, \gamma_{\rm V} 
    [(\bar\psi\gamma_\mu\psi)(\bar\psi\gamma^\mu\psi)]^2,
\\[6pt]
  {\cal L}^{\rm der} 
  & = & 
  - \tfrac{1}{2}\,\delta_{\rm S}(\partial_\nu\bar\psi\psi)
    (\partial^\nu\bar\psi\psi)  
  - \tfrac{1}{2}\,  \delta_{\rm V} (\partial_\nu\bar\psi\gamma_\mu\psi)
    (\partial^\nu\bar\psi\gamma^\mu\psi)
\\[6pt]
  & &
   - \tfrac{1}{2}\, \delta_{\rm TS} (\partial_\nu\bar\psi\vec\tau\psi) \cdot
   (\partial^\nu\bar\psi\vec\tau\psi)
  - \tfrac{1}{2}\, \delta_{\rm TV} (\partial_\nu\bar\psi\vec\tau\gamma_\mu\psi)
    \cdot (\partial^\nu\bar\psi\vec\tau\gamma^\mu\psi),
\\[6pt]
  {\cal L}^{\rm em} 
  & = & 
  -  e A_\mu\bar\psi[(1-\tau_3)/2]\gamma^\mu\psi -  
    \tfrac{1}{4}\, F_{\mu\nu} F^{\mu\nu}.
\end{array}
\label{lagrangian}
\end{equation}
Vectors in isospin space are denoted by arrows, and bold-faced
symbols will indicate vectors in ordinary three-dimensional space.
In addition to the free nucleon Lagrangian $\mathcal{L}_{\rm free}$,
the four-fermion interaction terms contained in
$\mathcal{L}_{\rm 4f}$, and higher order terms in $\mathcal{L}_{\rm hot}$, 
when applied to finite nuclei the model
must include the coupling $\mathcal{L}_{\rm em}$
of the protons to the electromagnetic field $A^\mu$,
and derivative terms contained in $\mathcal{L}_{\rm der}$.
In the terms $\partial_\nu(\bar\psi \Gamma \psi)$ the derivative is
understood to act on both $\bar\psi$ and $\psi$. 
One could, of course, construct many more higher order interaction terms, 
or derivative terms of higher order, but in practice only a relatively 
small set of free parameters can be adjusted from the data set of 
ground state nuclear properties.

The single-nucleon Dirac equation is derived from the variation of the
Lagrangian (\ref{lagrangian}) with respect to $\bar{\psi}$
\begin{equation}
\left\{ \bm{\alpha}\left[ -i\bm{\nabla}-\bm{V}({\bm r}) \right] +V({\bm r})+
\beta \big(m+S({\bm r})\big) \right\} \psi_{i}({\bm r}) = \epsilon_i\psi_{i}({\bm r})\;.
\label{dirac}
\end{equation}
The scalar and vector potentials
\begin{equation}
S({\bm r}) = \Sigma_S({\bm r}) + \tau_3\Sigma_{TS}({\bm r})\;,
\label{scapot}
\end{equation}
\begin{equation}
V^{\mu}({\bm r})  = \Sigma^{\mu}({\bm r}) + \tau_3\Sigma^{\mu}_{TV}({\bm r})\;,
\label{vecpot}
\end{equation}
contain the nucleon isoscalar-scalar, isovector-scalar, 
isoscalar-vector and isovector-vector self-energies 
defined by the following relations:
\begin{eqnarray}
   \label{selfS}
   \Sigma_S & = & \alpha_S \rho_S + \beta_S \rho_S^2 + 
   \gamma_S\rho_S^3+ \delta_S \triangle \rho_S \; ,\\
   \label{selfTS}
   \Sigma_{TS} & = & \alpha_{TS} \rho_{TS}
   +\delta_{TS} \triangle \rho_{TS}\; , \\
   \label{selfV}
   \Sigma^{\mu} & = & \alpha_V j^{\mu} 
  +\gamma_V (j_\nu j^\nu)j^\mu + \delta_V \triangle j^\mu     
   -eA^\mu\frac{1-\tau_3}{2} \; ,\\
   \label{selfTV}
   \Sigma^{\mu}_{TV} & = & \alpha_{TV} j^{\mu}_{TV} 
      + \delta_{TV} \triangle j^{\mu}_{TV}\;,      
\end{eqnarray}
respectively. Because of charge conservation, 
only the $3-rd$ component of the isovector
densities and currents contributes to the nucleon self-energies.
The local densities and currents are calculated in 
the {\it no-sea} approximation
\begin{eqnarray}
\label{dens_1}
\rho_{S}({\bm r}) &=&\sum\limits_{i=1}^{A}\psi_{i}^{\dagger}({\bm r})
             \beta\psi _{i}({\bm r})~,  \\
\label{dens_2}
\rho_{TS}({\bm r}) &=&\sum\limits_{i=1}^{A}
      \psi_{i}^{\dagger}({\bm r})\beta\tau_3\psi _{i}^{{}}({\bm r})~,  \\
\label{dens_3}
j^{\mu}({\bm r}) &=&\sum\limits_{i=1}^{A}\psi_{i}^{\dagger}({\bm r})
        \beta\gamma^\mu\psi _{i}^{{}}({\bm r})~,  \\
\label{dens_4}
j^{\mu}_{TV}({\bm r}) &=&\sum\limits_{i=1}^{A}\psi_{i}^{\dagger}({\bm r})
     \beta\gamma^\mu \tau_3 \psi _{i}^{{}}({\bm r})~.  
\end{eqnarray}
For a nucleus with A nucleons, the
summation runs over all occupied states in the Fermi sea, i.e. only
occupied single-nucleon states with positive energy explicitly contribute to 
the nucleon self-energies.
The energy momentum tensor determines the total energy of the nuclear system
\begin{eqnarray}
{{E}}_{RMF} &=& \int d{\bm r }~{\mathcal{E}_{RMF}}(\bm{r}) \nonumber \\ &=&
\sum_i{\int d\bm{r}~{\psi_i^\dagger (\bm{r}) \left( -i\bm{\alpha}
 \bm{\nabla} + \beta m\right )\psi_i(\bm{r})}} \nonumber \\
 &+& \int d{\bm r }~{\left (\frac{\alpha_S}{2}\rho_S^2+\frac{\beta_S}{3}\rho_S^3 +
  \frac{\gamma_S}{4}\rho_S^4+\frac{\delta_S}{2}\rho_S\triangle \rho_S
 + \frac{\alpha_V}{2}j_\mu j^\mu + \frac{\gamma_V}{4}(j_\mu j^\mu)^2 +
       \frac{\delta_V}{2}j_\mu\triangle j^\mu \right.} \nonumber \\
 &+& \left .
  \frac{\alpha_{TV}}{2}j^{\mu}_{TV}(j_{TV})_\mu+\frac{\delta_{TV}}{2}
    j^\mu_{TV}\triangle  (j_{TV})_{\mu}
 + \frac{\alpha_{TS}}{2}\rho_{TS}^2+\frac{\delta_{TS}}{2}\rho_{TS}\triangle 
      \rho_{TS} +\frac{e}{2}\rho_p A^0
 \right) ,
\label{EMF}
\end{eqnarray}
where $\rho_p$ denotes the proton density, and $A^0$ is the Coulomb potential.

In this work we only consider even-even nuclei that can be described 
by axially symmetric shapes. It is therefore convenient to work in
cylindrical coordinates
\begin{equation}
x=r_\perp \cos{\phi}, \qquad y=r_\perp \sin{\phi} \qquad \textrm{and} \qquad z\;.
\label{cylcoor} 
\end{equation}
In addition to axial symmetry, parity, symmetry with respect to the 
operator $e^{-i\pi\hat{J}_y}$, and time-reversal invariance are 
imposed as self-consistent symmetries. 
Time-reversal invariance implies that spatial components of the currents
vanish in the nuclear ground state. The resulting single-nucleon 
Dirac equation reads 
\begin{equation}
\left\{ -i\bm{\alpha}\bm{\nabla} +V({\bm r})+
\beta \big(m+S({\bm r})\big) \right\} \psi_{i}({\bm r}) = 
\epsilon_i\psi_{i}({\bm r})\;.
\label{dirac2}
\end{equation}
The eigensolutions are characterized by the
projection of the total angular momentum along the symmetry axis ($\Omega_i$),
the parity ($\pi_i$), and the $z$-component of the isospin ($t_i$). The Dirac
spinor has the following form
\begin{equation}
\psi_i(\bm{r},t) = \left( \begin{array}{c}
                f_i(\bm{r},s,t) \\
		ig_i(\bm{r},s,t) \end{array} \right) = \frac{1}{\sqrt{2\pi}}
		\left( \begin{array}{c}
		f_i^+(z,r_\perp )e^{i(\Omega_i -1/2)\phi} \\
		f_i^-(z,r_\perp )e^{i(\Omega_i +1/2)\phi} \\
		ig_i^+(z,r_\perp )e^{i(\Omega_i -1/2)\phi} \\
		ig_i^-(z,r_\perp )e^{i(\Omega_i +1/2)\phi} \end{array} \right)
		\chi_{t_i}(t) \;.		
\label{spinor}
\end{equation}
For each solution with positive $\Omega$
\begin{equation}
\psi_i \equiv \{ f_i^+,f_i^-,g_i^+,g_i^-;\Omega_i \} \;,
\label{omegapos}
\end{equation}
the corresponding degenerate time-reversed state
\begin{equation}
\psi_{~\bar{i}} = T\psi_i = \{ -f_i^-,f_i^+,g_i^-,-g_i^+;-\Omega_i \} \;,
\label{omeganeg}
\end{equation}
is obtained by acting with the time-reversal operator $T=i\sigma_yK$.
For even-even nuclei, the time-reversed states $i$ and $\bar{i}$
have identical occupation probabilities. 

The single-nucleon Dirac eigenvalue equation is solved by expanding the
spinors $f_i$ and $g_i$ Eq. (\ref{spinor}) in terms of eigenfunctions
of an axially symmetric harmonic oscillator potential~\cite{GTR.90}
\begin{equation}
V_{osc}(z,r_\perp ) = \frac{1}{2}M\omega_z^2z^2 + \frac{1}{2}M\omega_{\perp}^2
     r_\perp^2 \;.
\label{hopot}     
\end{equation}
Imposing volume conservation, the two
oscillator frequencies $\hbar \omega_z$ and $\hbar \omega_\perp$ can be
expressed in terms of the deformation parameter $\beta_0$ and the oscillator
frequency $\hbar \omega_0$
\begin{equation}
\hbar \omega_z = \hbar \omega_0 e^{-\sqrt{5/4\pi}\beta_0} 
\qquad \textrm{and} \qquad
\hbar \omega_\perp = \hbar \omega_0 e^{\frac{1}{2}\sqrt{5/4\pi}\beta_0} \;.
\label{frequencies}
\end{equation}
The corresponding oscillator length parameters are 
$\displaystyle b_z=\sqrt{\hbar/M\omega_z}$ and 
$\displaystyle b_\perp=\sqrt{\hbar/M\omega_\perp}$.
Because of volume conservation, $b_{{\perp }}^{2}b_{z}=b_{0}^{3}$.
The basis is now specified by the two constants $\hbar \omega _{0}$ and 
$\beta _{0}$, and basis states are characterized 
by the set of quantum numbers 
\begin{equation}
| \alpha > = | n_{z}, n_{\perp},\Lambda, m_{s}> \;,
\label{quantum}
\end{equation}
where $n_z$ and $n_{\perp}$ denote the number of nodes
in the $z$ and $r_\perp$ directions, respectively. 
$\Lambda$ and $m_{s}$ are the components of the orbital angular momentum
and the spin along the symmetry axis. 
The eigenvalue of $j_{z}$ -- the $z$-projection of the total single-nucleon 
angular momentum 
\begin{equation}
\Omega  = \Lambda+m_{s},
\end{equation}
and the parity is given by  
\begin{equation}
\pi = (-1)^{n_z+\Lambda}\;.
\label{parity}
\end{equation}

The eigenfunctions of the axially symmetric harmonic oscillator potential read
\begin{equation}
\Phi_\alpha(\bm{r},s) =
\frac{N_{n_z}}{\sqrt{b_z}}H_{n_z}(\xi)e^{-\xi^2/2}
\frac{N_{n_\perp}^{\Lambda}}{b_\perp}\sqrt{2}\eta^{\Lambda/2}
  L_{n_\perp}^\Lambda(\eta)e^{-\eta /2}
\frac{1}{2\pi}e^{i\Lambda\phi}\chi_{m_s}(s)\;,
\label{hoeig}
\end{equation}
with $\xi=z/b_z$ and $\eta=r_\perp^2/b_\perp^2$. The Hermite polynomials
$H_n(\xi)$, and the associated Laguerre polynomials $L_n^\Lambda(\eta)$, are
defined in Ref.~\cite{AS.65}. The normalization factors are given by
\begin{equation}
N_{n_z}=\frac{1}{\sqrt{\sqrt{\pi}2^{n_z}n_z!}} \qquad \textrm{and} \qquad
  N_{n_\perp}^\Lambda=\sqrt{\frac{n_\perp!}{(n_\perp+1)!}}\;.
\label{hoeignorm}  
\end{equation}
The large and small components of the single-nucleon Dirac spinor Eq. 
(\ref{spinor}) are expanded in terms of the eigenfunctions Eq. (\ref{hoeig}) 
\begin{equation}
f_i(\bm{r},s,t) = \frac{1}{\sqrt{2\pi}} \left(
\begin{array}{c}
f_i^+(z,r_\perp)e^{i(\Omega-1/2)\phi} \\
f_i^-(z,r_\perp)e^{i(\Omega+1/2)\phi} 
\end{array} \right)
=\sum_{\alpha}^{\alpha_{max}}f_i^\alpha \Phi_\alpha(\bm{r},s) \chi_{t_i}(t) \;,
\label{f_exp}
\end{equation}
\begin{equation}
g_i(\bm{r},s,t) = \frac{1}{\sqrt{2\pi}} \left(
\begin{array}{c}
g_i^+(z,r_\perp)e^{i(\Omega-1/2)\phi} \\
g_i^-(z,r_\perp)e^{i(\Omega+1/2)\phi} 
\end{array} \right)
=\sum_{\tilde{\alpha}}^{\tilde{\alpha}_{max}}g_i^{\tilde{\alpha}} 
    \Phi_{\tilde{\alpha}}(\bm{r},s) \chi_{t_i}(t) \;.
\label{g_exp}    
\end{equation}
In order to avoid the onset of spurious states, the quantum
numbers $\tilde{\alpha}_{max}$ and $\alpha_{max}$ are chosen in such a
way that the corresponding major oscillator quantum numbers 
$N=n_z+2n_{\perp}+\Lambda$ are not larger than $N_{sh}+1$ for 
the expansion of the small components, and not larger than $N_{sh}$ for 
the expansion of the large components \cite{GTR.90}. 

For an axially deformed nucleus the map of the energy surface 
as a function of the quadrupole moment is obtained by imposing a
constraint on the mass quadrupole moment. The method of quadratic 
constraint uses an unrestricted variation of the function
\begin{equation}
<H>~+~\frac{C}{2}\left( <\hat{Q} > - ~q \right)^2 \;,
\label{constr}
\end{equation} 
where $<H>$ is the total energy, $<\hat{Q}>$ denotes the expectation 
value of the mass quadrupole operator, $q$ is the deformation parameter, 
and $C$ is the stiffness constant~\cite{RS.80}.

In addition to the self-consistent mean-field potential, for open-shell 
nuclei pairing correlations have to be included in the energy functional.
In this work we do not consider nuclear systems very far from the valley of 
$\beta$-stability, and therefore a good approximation for the treatment
of pairing correlations is provided by the BCS formalism. 
Following the prescription from Ref.~\cite{BMM.02}, we use a
$\delta$ force in the pairing channel, supplemented with a smooth 
cut-off determined by a Fermi function in the single-particle energies. 
The pairing contribution to the total energy is given by
\begin{equation}
{E}_{pair}^{p(n)} = \int{\mathcal{E}_{pair}^{p(n)}(\bm{r})d\bm{r}}= 
   \frac{V_{p(n)}}{4}\int{\kappa_{p(n)}^*(\bm{r})\kappa_{p(n)}(\bm{r}) d\bm{r}}\;,
\label{epair}	    
\end{equation}
for protons and neutrons, respectively. $\kappa_{p(n)}(\bm{r})$ denotes 
the local part of the pairing tensor, and $V_{p(n)}$ is the pairing strength
parameter. Of course, for open-shell nuclei the expressions 
Eqs. (\ref{dens_1}) -- (\ref{dens_4}) for the local densities and currents 
include the occupation factors of single-nucleon states. 
Finally, the expression for the total energy reads
\begin{equation}
E_{tot} =\int{\left[\mathcal{E}_{RMF}(\bm{r})+\mathcal{E}_{pair}^p(\bm{r})
         +\mathcal{E}_{pair}^n(\bm{r})\right]d\bm{r} }\;.
\label{etot}
\end{equation}
The center-of-mass correction has been included by adding the expectation
value 
\begin{equation}
 E_{cm} = -\frac{\langle \hat{P}_{cm}^2 \rangle}{2mA}\;, 
\end{equation} 
to the total energy. $P_{cm}$ is the total momentum of a 
nucleus with $A$ nucleons.
\subsection{\label{subIIb}The generator coordinate method}	
The generator coordinate method (GCM) is based on the assumption that, starting
from a set of mean-field states $\ket{\phi (q)}$ which depend on a collective
coordinate $q$, one can build approximate eigenstates of the nuclear Hamiltonian 
\begin{equation}
\ket{\Psi_\alpha} = \sum_j{f_\alpha(q_j)\ket{\phi (q_j)}}\;.
\label{GCM-state}
\end{equation}
A detailed review of the GCM can be found in Chapter 10 of Ref.~\cite{RS.80}.
In this work the basis states $\ket{\phi (q)}$ are Slater 
determinants of single-nucleon states generated by solving the 
constrained relativistic mean-field + BCS equations, as described in the
previous section. This means that we use the mass quadrupole moment as the 
generating coordinate $q$. The axially deformed mean-field breaks rotational
symmetry, so that the basis states $\ket{\phi (q)}$ are not eigenstates of the
total angular momentum. Of course, in order to be able to compare 
theoretical predictions with data, it is necessary to construct states 
with good angular momentum
\begin{equation}
\ket{\Psi_\alpha^{JM}} = \sum_{j,K}{f_\alpha^{JK}(q_j)\hat{P}_{MK}^J
    \ket{\phi (q_j)}}\;,
\label{AMPGCM-state}
\end{equation}
where $\hat{P}^J_{MK}$ denotes the angular momentum projection operator
\begin{equation}
\hat{P}^J_{MK} = \frac{2J+1}{8\pi^2}\int{d\Omega D_{MK}^{J*}(\Omega )\hat{R}
   (\Omega )}\;.
\label{proj_op}   
\end{equation}
Integration is performed over the three Euler angles
$\alpha$, $\beta$, and $\gamma$.
$D_{MK}^{J}(\Omega )=e^{-iM\alpha}d^J_{MK}(\beta)e^{-iK\gamma}$ is the Wigner 
function~\cite{Var.88}, and 
$\hat{R}(\Omega )=e^{-i\alpha\hat{J}_z}e^{-i\beta\hat{J}_y}
e^{-i\gamma\hat{J}_z}$ is the rotation operator.
The weight functions $f_{\alpha}^{JK}(q_j)$ are determined from a variational
calculation, 
\begin{equation} 
 \delta E^{J} =
 \delta \frac{\bra{\Psi_\alpha^{JM}} \hat{H} \ket{\Psi_\alpha^{JM}}}
            {\bra{\Psi_\alpha^{JM}}\Psi_\alpha^{JM}\rangle} = 0 \; ,
\label{variational}
\end{equation}
i.e. by requiring that the expectation value of the energy is stationary
with respect to an arbitrary variation $\delta f_{\alpha}^{JK}$. This 
leads to the Hill-Wheeler equation
\begin{equation}
\sum_{j,K}f_{\alpha}^{JK}(q_j)
  \left( \left\langle\phi(q_i) \right|\hat{H}\hat{P}_{MK}^J\left|
  \phi(q_j)\right\rangle - E^J_\alpha 
  \left\langle\phi(q_i) \right|\hat{P}_{MK}^J\left|\phi(q_j)\right\rangle \right) = 0\;.
\label{HWEQ}
\end{equation}
The restriction to 
axially symmetric configurations ($\hat{J}_z\ket{\phi(q)} = 0$)
simplifies the problem considerably, because in this case the 
integrals over the Euler angles $\alpha$ and $\gamma$ can be
performed analytically. For an arbitrary multipole operator 
$\hat{Q}_{\lambda \mu}$ one thus finds
\begin{equation}
\bra{\phi(q_i)}\hat{Q}_{\lambda \mu} \hat{P}_{MK}^J\ket{\phi(q_j)} =
\frac{2J+1}{2}\delta_{M-\mu}\delta_{K0}\int_0^\pi{\sin{\beta}d_{-\mu 0}^{J*}(\beta)
  \bra{\phi(q_i)} \hat{Q}_{\lambda \mu}
  e^{-i\beta\hat{J}_y}\ket{\phi(q_j)}d\beta}\;.
\label{matel}  
\end{equation}
By using the identity 
$e^{i\beta\hat{J}_y}=e^{-i\pi\hat{J}_z}e^{-i\beta\hat{J}_y} e^{i\pi\hat{J}_z}$,
together with parity, and the symmetry with respect to the operator 
$e^{-i\pi\hat{J}_y}$, the integration interval in Eq. (\ref{matel})
can be reduced from $[0,\pi]$  to $[0,\pi/2]$
\begin{eqnarray}
\bra{\phi(q_i)}\hat{Q}_{\lambda \mu} \hat{P}_{MK}^J\ket{\phi(q_j)} &=&
(2J+1)\frac{1+(-1)^J}{2}\delta_{M-\mu}\delta_{K0} \nonumber \\
 && \int_0^{\pi/2}{\sin{\beta}d_{-\mu 0}^{J*}(\beta) \bra{\phi(q_i)} 
  \hat{Q}_{\lambda \mu}e^{-i\beta\hat{J}_y}\ket{\phi(q_j)}d\beta}\;.
\label{matel2}  
\end{eqnarray}
We notice that this expression vanishes for odd values of angular momentum
$J$, i.e., the projected quantities are defined only for even values of $J$.

The norm overlap kernel
\begin{eqnarray}
\mathcal{N}^J(q_i,q_j) &=& \bra{\phi(q_i)} \hat{P}_{MK}^J\ket{\phi(q_j)} =
\nonumber \\ &&
(2J+1)\frac{1+(-1)^J}{2}\delta_{M0}\delta_{K0} 
  \int_0^{\pi/2}{\sin{\beta}d_{00}^{J*}(\beta) \bra{\phi(q_i)} 
  e^{-i\beta\hat{J}_y}\ket{\phi(q_j)}d\beta}\;,
\label{normker}  
\end{eqnarray}
can be evaluated by employing
the generalized Wick theorem~\cite{OY.66,BB.69,Bon.90,VHB.00}
\begin{equation}
n(q_i,q_j;\beta) \equiv \bra{\phi(q_i)} e^{-i\beta\hat{J}_y}\ket{\phi(q_j)}
  = \pm\sqrt{det~\mathcal{N}_{ab}(q_i,q_j;\beta)}\;.
\label{norm}
\end{equation}
The overlap matrix is defined:
\begin{equation}
\mathcal{N}_{ab}(q_i,q_j;\beta) = u_a(q_i)R_{ab}(q_i,q_j;\beta)u_b(q_j) +
                         v_a(q_i)R_{ab}(q_i,q_j;\beta)v_b(q_j) \;,
\label{N_mat}
\end{equation}
where $u$ and $v$ denote the BCS occupation probabilities, and the matrix $R$ 
reads
\begin{equation}
R_{ab}(q_i,q_j;\beta) = \int{\psi_a^\dagger(\bm{r};q_i)e^{-i\beta\hat{J}_y}
                     \psi_b(\bm{r};q_j)d\bm{r}}\;.
\label{R_mat}
\end{equation}
If the expansions Eqs. (\ref{f_exp}) and (\ref{g_exp}) are 
inserted in the expression above, the evaluation of
the matrix $R$ reduces to the calculation of the matrix elements of the
rotation operator in the basis of the axially symmetric harmonic oscillator
\begin{eqnarray}
R_{ab}(q_i,q_j) &=& \sum_{\alpha,\beta}{f_a^{\alpha}(q_i)f_b^{\beta}(q_j)
                 \left< \alpha \left| e^{-i\beta\hat{J}_y} \right| \beta
		  \right> } \nonumber \\
	       &+& \sum_{\tilde{\alpha},\tilde{\beta}}{g_a^{\tilde{\alpha}}(q_i)
	           g_b^{\tilde{\beta}}(q_j)\left< \tilde{\alpha} \left|
		    e^{-i\beta\hat{J}_y} \right|\tilde{\beta}\right> }\;. 
\label{mat_Rosc}		    
\end{eqnarray}
The simplest way to evaluate these matrix elements is to express
the eigenfunctions of the axially symmetric harmonic oscillator
in the spherically symmetric oscillator basis. The transformation
from the spherical to the axially deformed basis is given by 
the following expression
\begin{equation}
\ket{\Omega_\alpha \Lambda_\alpha n_{\perp}^{\alpha} n_z^{\alpha}}
= \sum_{nlj}{S_{
   \Omega_\alpha \Lambda_\alpha n_{\perp}^{\alpha} n_z^{\alpha}}^{nlj}}
  \ket{nlj\Omega_\alpha} \;,
\label{cyl2sph}  
\end{equation}
with the transformation coefficients 
$S_{\Omega_\alpha \Lambda_\alpha n_{\perp}^{\alpha} n_z^{\alpha}}^{nlj}$
given in Ref.~\cite{Tal.70}. It must be emphasized that this transformation
is only possible if $\omega_z = \omega_\perp$ (i.e. $\beta_0=0$ in 
Eq.~(\ref{frequencies}) )~\cite{ERS.93,Rob.94}. 
In addition, the same 
oscillator frequency $\hbar\omega_0$ has to be used for each
value of the generating coordinate $q$ in order to avoid completeness problems
in the GCM calculations~\cite{Rob.94}. We have used $\hbar\omega_0=41A^{-1/3}$.
Since the choice of the two basis parameters $\hbar \omega_0$ and $\beta_0$
cannot be optimized, the convergence of the results should be
carefully checked as a function of the number of oscillator shells
used in the expansion of the Dirac spinors. The expression for the 
matrix elements is simply
\begin{equation}
\bra{\alpha}e^{-i\beta\hat{J}_y}\ket{\beta} = \sum_{nlj}{
S_{\Omega_\alpha \Lambda_\alpha n_{\perp}^{\alpha}n_z^{\alpha}}^{nlj}
S_{\Omega_\beta \Lambda_\beta n_{\perp}^{\beta}n_z^{\beta}}^{nlj}
d_{\Omega_\alpha \Omega_\beta}^j(\beta)}\;,
\label{rot_mat_el}
\end{equation}
where $d_{\Omega_\alpha \Omega_\beta}^j(\beta)$ denotes the Wigner
rotation matrix~\cite{Var.88}.
More general transformation coefficients for the case 
$\omega_\perp \neq \omega_z$
have been derived in Ref.~\cite{Naz.96}, but they are rather complicated
and have not been used in the present analysis.

The Hamiltonian kernel 
\begin{eqnarray}
\mathcal{H}^J(q_i,q_j) &=& \bra{\phi(q_i)}\hat{H} \hat{P}_{MK}^J\ket{\phi(q_j)} =
\nonumber \\ &&
(2J+1)\frac{1+(-1)^J}{2}\delta_{M0}\delta_{K0} 
  \int_0^{\pi/2}{\sin{\beta}d_{00}^{J*}(\beta) \bra{\phi(q_i)}\hat{H} 
  e^{-i\beta\hat{J}_y}\ket{\phi(q_j)}d\beta}\;,
\label{hamker}  
\end{eqnarray}
can be calculated from the mean-field energy functional Eq. 
(\ref{EMF})~\cite{OY.66,BB.69,Bon.90,VHB.00}, provided the 
modified densities 
\begin{eqnarray}
\tau(\bm{r};q_i,q_j,\beta) &=&\sum_{a,b}v_a(q_i)v_b(q_j)
      \mathcal{N}_{ba}^{-1}(q_i,q_j;\beta)
      \psi_a^\dagger(\bm{r};q_i)(-i\bm{\alpha}\bm{\nabla} + m\beta)
      e^{-i\beta\hat{J}_y}\psi_b(\bm{r};q_j)\;, \\
\rho_{S}(\bm{r};q_i,q_j,\beta) &=&\sum_{a,b}v_a(q_i)v_b(q_j)
      \mathcal{N}_{ba}^{-1}(q_i,q_j;\beta)
      \psi_a^\dagger(\bm{r};q_i) \beta
      e^{-i\beta\hat{J}_y}\psi_b(\bm{r};q_j) \;, \\
\rho_{TS}(\bm{r};q_i,q_j,\beta) &=&\sum_{a,b}v_a(q_i)v_b(q_j)
      \mathcal{N}_{ba}^{-1}(q_i,q_j;\beta)
      \psi_a^\dagger(\bm{r};q_i) \beta \tau_3
      e^{-i\beta\hat{J}_y}\psi_b(\bm{r};q_j) \;, \\
j^\mu(\bm{r};q_i,q_j,\beta) &=&\sum_{a,b}v_a(q_i)v_b(q_j)
      \mathcal{N}_{ba}^{-1}(q_i,q_j;\beta)
      \psi_a^\dagger(\bm{r};q_i)\beta\gamma^\mu 
      e^{-i\beta\hat{J}_y}\psi_b(\bm{r};q_j) \;, \\
j^\mu_{TV}(\bm{r};q_i,q_j,\beta) &=&\sum_{a,b}v_a(q_i)v_b(q_j)
   \mathcal{N}_{ba}^{-1}(q_i,q_j;\beta) \psi_a^\dagger(\bm{r};q_i)
      \beta\gamma^\mu  \tau_3 e^{-i\beta\hat{J}_y}\psi_b(\bm{r};q_j) \;, 
\label{dens_gcm}
\end{eqnarray}
are used when evaluating the expression
\begin{equation}
h(q_i,q_j;\beta)\equiv \bra{\phi(q_i)}\hat{H} e^{-i\beta\hat{J}_y}\ket{\phi(q_j)}
  =\int{\mathcal{E}_{tot}(\bm{r};q_i,q_j,\beta)d\bm{r}}\;.
\end{equation}
The computational task of evaluating the Hamiltonian and norm overlap kernels
can be reduced significantly if one realizes that states with very small
occupation probabilities give negligible contributions to the kernels.
Such states can be excluded from the calculation, and the details of
this procedure can be found in Refs.~\cite{Bon.90,VHB.00}.

An additional problem arises from the fact that the basis 
states $\ket{\phi(q_j)}$ are not eigenstates of the proton
and neutron number operators $\hat{Z}$ and $\hat{N}$. The adjustment of the
Fermi energies in a BCS calculation ensures only that the average value of
the nucleon number operators corresponds to the actual number of nucleons.
Consequently, the wave functions $\ket{\Psi_\alpha^{JM}}$ are generally not
eigenstates of the nucleon number operators and, moreover, the average values
of the nucleon number operators are not necessarily equal to the number of
nucleons in a given nucleus. This happens because the binding energy increases
with the average number of nucleons and, therfore, an unconstrained 
variation of the weight functions in a GCM calculation will generate 
a ground state with the average number of protons and neutrons larger
than the actual values in a given nucleus. In order to restore the 
correct mean values of the nucleon numbers, we follow the usual 
prescription~\cite{Bon.90,HHR.82} and modify the Hill-Wheeler equation 
by replacing $h(q_i,q_j;\beta)$ with
\begin{equation}
h^\prime(q_i,q_j;\beta) = h(q_i,q_j;\beta)
 -\lambda_p\left[ z(q_i,q_j;\beta)-z_0\right]
  -\lambda_n\left[ n(q_i,q_j;\beta)-n_0\right]\;,
\end{equation}
where
\begin{equation}
z(q_i,q_j;\beta)=\bra{\phi(q_i)}\hat{Z} e^{-i\beta\hat{J}_y}\ket{\phi(q_j)}
\quad \textrm{and} \quad
n(q_i,q_j;\beta)=\bra{\phi(q_i)}\hat{N} e^{-i\beta\hat{J}_y}\ket{\phi(q_j)}\;.
\end{equation}
$\lambda_{p(n)}$ is the proton (neutron) Fermi energy, while $z_0$ and 
$n_0$ denote the desired number of protons and neutrons, respectively.

The Hill-Wheeler equation 
\begin{equation}
\sum_j{\mathcal{H}^J(q_i,q_j)f^J_\alpha(q_j)} = E^J_\alpha
        \sum_j{\mathcal{N}^J(q_i,q_j)f^J_\alpha(q_j)}\;,
\label{HWEQ2}	
\end{equation}
presents a generalized eigenvalue problem, and thus the weight functions 
$f^J_\alpha(q_i)$ are not orthogonal and cannot be interpreted as collective
wave functions for the variable $q$. It is useful to re-express
Eq. (\ref{HWEQ2}) in terms of another set of functions,  $g^J_\alpha(q_i)$,
defined by 
\begin{equation}
g^J_\alpha(q_i) = \sum_j(\mathcal{N}^{J})^{1/2}(q_i,q_j) f^J_\alpha(q_j)\;.
\label{coll_wf}
\end{equation}
With this transformation the
Hill-Wheeler equation defines an ordinary eigenvalue problem
\begin{equation}
\sum_j{\tilde{\mathcal{H}}^J(q_i,q_j)g^J_\alpha(q_j)} =E_\alpha g_\alpha^J(q_i)\;,
\end{equation}
with
\begin{equation}
\tilde{\mathcal{H}}^J(q_i,q_j) = \sum_{k,l}(\mathcal{N}^{J})^{-1/2}(q_i,q_k) 
  \mathcal{H}^J(q_k,q_l) (\mathcal{N}^{J})^{-1/2}(q_l,q_j) \;.
\end{equation}
The functions $g^J_\alpha(q_i)$ are orthonormal and play the role 
of collective wave functions.

In practice, the first step in the solution of Eq. (\ref{HWEQ2}) is the
diagonalization of the 
norm overlap kernel $\mathcal{N}^J(q_i,q_j)$ Eq.~(\ref{normker})
\begin{equation}
\sum_j{\mathcal{N}^J(q_i,q_j)u_k(q_j)} = n_ku_k(q_i)\;.
\label{diag_norm}
\end{equation}
Since the basis functions 
$\ket{\phi (q_i)}$ are not linearly independent, many of the 
eigenvalues $n_k$ are very close to zero. The corresponding eigenfunctions
$u_k(q_i)$ are rapidly oscillating and carry very little physical information.
However, due to numerical uncertainties, their contribution to 
$\tilde{\mathcal{H}}^J(q_i,q_j)$ can be large, and these states should be
removed from the basis. From the remaining states one builds the collective
Hamiltonian
\begin{equation}
\mathcal{H}^{Jc}_{kl} = \frac{1}{\sqrt{n_k}}\frac{1}{\sqrt{n_l}}
  \sum_{i,j}{u_k(q_i)\tilde{\mathcal{H}}^J(q_i,q_j)u_l(q_j)}\;,
\label{Hcoll}  
\end{equation}
which is subsequently diagonalized
\begin{equation}
\sum_{k,l}\mathcal{H}^{Jc}_{kl}g_l^{J\alpha} = E^J_{\alpha}g_k^{J\alpha} \;.
\label{mat_coll}
\end{equation}
The solution determines both the ground state energy, and the energies 
of excited states, for each value of the angular momentum $J$. 
The collective wave functions $g_\alpha^J(q)$, and the weight functions
$f_\alpha^J(q)$, are calculated from the norm overlap eigenfunctions
\begin{equation}
g_\alpha^J(q_i) = \sum_l{ g_l^{J\alpha} u_l(q_i) }\;,
\label{g_u}
\end{equation}
and 
\begin{equation}
f_\alpha^J(q_i) = \sum_l{\frac{g_l^{J\alpha}}{\sqrt{n_l}} u_l(q_i)} \;.
\label{f_u}
\end{equation}

Once the weight functions $f^J_\alpha(q)$ are known, it is
straightforward to calculate all physical observables, such as transition
probabilities and spectroscopic quadrupole moments~\cite{GER.02}.
The reduced transition probability for a transition between an initial state
$(J_i,\alpha_i)$, and a final state $(J_f,\alpha_f)$, reads
\begin{equation}
B(E2; J_i\alpha_i \to J_f\alpha_f) = \frac{e^2}{2J_i+1}\left| \sum_{q_f,q_i}{
f^{J_f*}_{\alpha_f}(q_f)\bra{J_fq_f}|\hat{Q}_2|\ket{J_iq_i}
f^{J_i}_{\alpha_i}(q_i)}\right|^2\;,
\label{BE2}
\end{equation}
and the spectroscopic quadrupole moment for a state $(J\alpha)$ is defined
\begin{equation}
Q^{spec}(J,\alpha)=e\sqrt{\frac{16\pi}{5}}
\left( \begin{array}{ccc}
J & 2 & J \\
J & 0 & -J \end{array} \right)
\sum_{q_i,q_j}{f^{J*}_{\alpha}(q_i)\bra{Jq_i}|\hat{Q}_2|\ket{Jq_j}
f^{J}_{\alpha}(q_j)}\;.
\label{Qspec}
\end{equation}
Since these quantities are calculated in full configuration space,
there is no need to introduce effective charges, hence $e$ denotes
the bare value of the proton charge. In order to evaluate transition
probabilities and spectroscopic quadrupole moments, we will need the reduced
matrix element of the quadrupole operator
\begin{eqnarray}
\bra{J_fq_f}|\hat{Q}_2|\ket{J_iq_i}&=&(2J_i+1)(2J_f+1)\sum_\mu
\left( \begin{array}{ccc}
J_i & 2 & J_f \\
-\mu & \mu & 0 \end{array} \right) \nonumber \\ &&
\int^{\pi/2}_0 {\sin{\beta}d_{-\mu0}^{J_i*}(\beta)\bra{\phi(q_f)}\hat{Q}_{2\mu}
 e^{-i\beta\hat{J}_y}\ket{\phi(q_i)}}\;.
\label{Q2red}
\end{eqnarray}
\section{\label{secIII}Illustrative calculations}

In this section we perform several illustrative configuration mixing 
calculations that will test our implementation of the generator 
coordinate method, as well as the angular momentum projection. 
The intrinsic wave functions that are used in the configuration 
mixing calculation have been obtained as solutions of the self-consistent
relativistic mean-field equations, subject to constraint on the 
mass quadrupole moment. The interaction in the particle-hole 
channel is determined by the effective point-coupling Lagrangian Eq. 
(\ref{lagrangian}), and a density-independent $\delta$-force is 
used as the effective interaction in the particle-particle channel.
Pairing correlations are treated within the BCS framework. 

Among a number of self-consistent RMF-PC models that have 
been considered over the last ten years, a few reliable and accurate 
phenomenological parameterizations have been adjusted and applied 
in the description of ground state properties of finite nuclei on
a quantitative level. In particular, based on an extensive  
multiparameter $\chi^2$ minimization procedure,  
B\"urvenich et al. have adjusted the PC-F1 set of coupling constants 
for an effective point-coupling Lagrangian with higher order interaction 
terms~\cite{BMM.02}. While the Lagrangian of Eq. (\ref{lagrangian}) contains 11 
adjustable coupling constants, the PC-F1 effective interaction
corresponds to a restricted set of 9 coupling parameters and
does not include the isovector-scalar channel.
In addition, the effective pairing interaction is determined by 
the strength parameters $V_p$ and $V_n$, for
protons and neutrons, respectively. The parameters in the particle-hole 
and particle-particle channels have been adjusted to ground state 
observables (binding energies, charge radii, diffraction radii, surface 
thickness, and pairing gaps) of 17 spherical nuclei~\cite{BMM.02}.

The PC-F1 interaction has been tested
in the analysis of the equations of state of symmetric nuclear matter and 
neutron matter, binding energies and form-factors, and shell-structure-related
ground-state properties of several isotopic and isotonic chains. This 
interaction has also been employed in relativistic quasiparticle random
phase approximation calculations of multipole giant 
resonances \cite{NVR.05}. A comparison with data has shown that the
RMF-PC model with the PC-F1 interaction can compete with the best
phenomenological finite-range meson-exchange interactions. It should be noted,
however, that PC-F1 exhibits a relatively large volume asymmetry at saturation,
resulting in a very stiff equation of state for neutron matter and too large
values for the neutron skin in finite nuclei. Modern meson-exchange
effective interactions, on the other hand, include an explicit medium 
dependence in both isoscalar and isovector channels~\cite{TW.99,NVFR.02,Lal.05}, 
and thus provide an improved description of asymmetric nuclear matter and 
neutron matter, and realistic values of the neutron skin in finite nuclei.

\subsection{\label{subIIIa}Test of the generator coordinate method: $^{194}$Hg}

Our first example is a test of the generator coordinate method in 
configuration mixing calculations for the nucleus $^{194}$Hg. At 
this stage we do not consider angular momentum projection yet. The 
results of the test for the ground and excited states will be directly
compared with the classical analysis of the GCM in the study of shape 
isomerism in $^{194}$Hg by Bonche {\it et al.} \cite{Bon.90}. It has to
be emphasized, however, that the calculated GCM energies cannot be 
compared with data on a quantitative level, because without 
angular momentum projection not only the rotational energy correction 
is missing, but also the overlaps between states which belong to prolate 
and oblate minima are significantly reduced \cite{BBH.04}. 

The GCM basis is constructed from self-consistent solution of the 
constrained single-nucleon Dirac equation on a regular mesh in the 
generating coordinate -- the mass quadrupole moment: 
from $q=-40$b to $q=80$b, with a spacing of $\Delta q=2$b. The GCM basis
thus consists of 61 intrinsic states. The large and small
components of Dirac spinors are expanded in terms of the axially symmetric
oscillator eigenfunctions. As already pointed out in Sec. \ref{subIIb},
the same oscillator frequency $\hbar\omega_0$ is used for each value of the
generating coordinate $q$, and additionally the condition 
$\omega_z = \omega_{\perp}$ is imposed. Since the basis parameters are fixed
to $\hbar\omega_0=41A^{-1/3}$ and $\beta_0=0$, rather then optimized,
the convergence of the results with respect to the number of major oscillator
shells used in the expansions Eqs. (\ref{f_exp}) and (\ref{g_exp}) has to be 
checked carefully.

In Fig.~\ref{figA} we display the binding energy curves for $^{194}$Hg,
as functions of the mass quadrupole moment, calculated by expanding the 
Dirac spinors in $10, 12, 14, 16$ and $18$ oscillator shells. Obviously, 
at least $14$ oscillator shells are necessary in order to obtain convergence
for deformations smaller than $q=35$b. Larger deformations require at 
least $16$ major oscillator shells. The absolute minimum of the 
binding energy curve corresponds to a slightly oblate shape ($q= -10$b). 
An additional shallow minimum at excitation energy $\approx 1.5$ MeV is 
found on the prolate side $q=6$ b. At much larger deformation, $q= 45$ b,
we find a third, superdeformed minimum $4.2$ MeV above the first minimum
of the binding energy curve. The deformations at which the three minima 
occur are in quantitative agreement with those calculated with the 
non-relativistic constrained Hartree-Fock plus BCS model of 
Ref.~\cite{Bon.90}, using the SIII Skyrme effective interaction. The 
excitation energy of the second minimum is $\approx 1.5$ MeV in both 
models, whereas the superdeformed minimum calculated with the SIII 
interaction is more than 2 MeV higher than in the present calculation. 
Unless stated otherwise, all calculations presented in this section
have been performed in the deformed oscillator basis with $N_{sh}=16$ 
oscillator shells.

The first step in the solution of the modified Hill-Wheeler equation is 
the construction and diagonalization of the norm overlap kernel
\begin{equation}
\mathcal{N}(q_i,q_j)=\langle \phi(q_i)|\phi(q_j) \rangle\;, 
\end{equation}
see Eq.~(\ref{diag_norm}).
Since the GCM basis states are not linearly independent, many of the 
norm overlap kernel eigenvalues $n_k$ are close to zero. 
This is illustrated in the left panel of Fig.~\ref{figB}, where we 
display the eigenvalues $n_k$ for four
different values of the mesh spacing, 
ranging from $\Delta q=2$b to $\Delta q=8$b.
For $\Delta q=2$b the overlaps between neighboring states 
are typically $\approx 0.8$, and in the corresponding set of $61$ eigenvalues
we find $13$ values smaller then $10^{-3}$. If the mesh spacing is
increased to $\Delta q=4$b and $\Delta q=6$b, the overlaps between 
neighboring states are reduced approximately by factors of two and eight, 
respectively. A further increase of the mesh spacing results in 
very small overlaps between neighboring states ($\approx 0.05$), 
i.e. basis states become almost orthogonal. Except in a few test cases that
will be specified explicitly, all calculations in this section have been
carried out with the mesh spacing $\Delta q=2$b. 

In the next step the GCM basis space is truncated by eliminating those
eigenvectors of the norm overlap kernel, which correspond to eigenvalues 
smaller than a given positive constant $\epsilon_n$. This is necessary 
in order to eliminate numerical instabilities in the diagonalization 
of the collective Hamiltonian Eq.~(\ref{Hcoll}). 
In the right panel of Fig.~\ref{figB}, the energies of twelve lowest
GCM states are plotted as functions of the number of basis states. We notice
that the spectrum is stable for a broad range of basis dimensions, between
25 ($\epsilon_n = 0.2$) and 55 ($\epsilon_n=5\cdot 10^{-4}$) vectors.
These results can be directly compared with Fig. 4 of Ref.~\cite{Bon.90}.
In the following calculations, eigenvectors of the norm overlap kernel with 
eigenvalues smaller than $\epsilon_n= 5\cdot 10^{-4}$ are eliminated from 
the basis.

In Fig.~\ref{figC} we plot the energies of fifteen lowest GCM states 
as functions of the average quadrupole moment 
\begin{equation}
\langle q_k \rangle = \sum_j{g_k^2(q_j) q_j}\;,
\label{avquad}
\end{equation}
calculated in oscillator bases with $10,12,14$ and $16$ oscillator shells, 
together with the corresponding mean-field binding energy curves.
The GCM ground states are normalized to zero energy. In all four cases 
the average deformation of the ground state is close to the minimum of
the binding energy curve, and the gain in correlation energy which 
results from configuration mixing is $\approx 0.8$ MeV. The energies
of the ground state and the two first excited states basically converge
already for a basis with $12$ shells. Higher excited states, however,
contain sizeable admixtures of basis states with larger deformations, and 
the corresponding energy spectrum is sensitive to the number of oscillator 
shells. 

The GCM states can be analyzed in more detail if one plots
their collective wave functions $g_k(q)$ as functions of the quadrupole moment.
In Fig.~\ref{figD} we display the collective wave functions for the first
fourteen GCM states in $^{194}$Hg. The vertical dashed line denotes the 
position of the barrier between the main potential well and the 
superdeformed well. Except for the fifth state, the wave functions of the 
lowest nine states are concentrated in the main potential well. The fifth
state obviously belongs to the superdeformed minimum, hence its energy 
displays a strong dependence on the number of oscillator shells (see also 
Fig.~\ref{figC}). For states with $k \ge 10$ the wave functions are  
generally spread over a wide region of deformations, both in the main 
and in the superdeformed well.

In Fig.~\ref{figE} we plot the GCM energy spectra, calculated with $16$ 
oscillator shells, for four values of the mesh spacing, ranging
from $\Delta q=2$b to $\Delta q=8$b. The corresponding mean-field 
binding energy curves are also included in the figure, and their 
minima are placed at zero energy. Comparing with our standard value of 
$\Delta q=2$b, we notice that the low-energy part
of the spectrum is accurately calculated also for $\Delta q=4$b. Increasing 
the mesh spacing to $\Delta q=6$b, accurate energies are obtained only for
the two lowest states. With a further increase of $\Delta q$, the overlaps 
between neighboring states become so small that there is hardly any 
configuration mixing. The resulting GCM energies are very close to the 
energies of the basis states $\ket{\phi(q)}$.

Several additional tests, carried out in comparison with the results of 
Ref.~\cite{Bon.90}, have shown that our implementation of the GCM is 
numerically stable and, therefore, it can also be used for configuration
mixing calculations with angular momentum projected states.

\subsection{\label{subIIIb}Test of angular momentum projection: $^{32}$Mg}

For a quantitative description of structure phenomena, especially in 
transitional deformed nuclei characterized by a coexistence of 
spherical and intruder configurations, calculations must explicitly include 
correlations related to restoration of broken symmetries. In particular, 
the rotational energy correction, i.e. the gain in energy obtained by 
projection on states with good angular momentum, can be
of the order of 2--4 MeV for the ground state. 
Here we perform several tests of the angular 
momentum projection for the isotope $^{32}$Mg. This nucleus belongs to 
the island of inversion at $N=20$, which is characterized by the 
melting of the neutron shell closure and the predominance of 
intruder state configurations in ground states of neutron-rich 
systems. The structure of $^{32}$Mg has been the subject of 
numerous experimental and theoretical studies. Several modern 
theoretical approaches have recently been employed in extensive 
studies of the erosion of the spherical $N=20$ shell closure in this 
neutron-rich nucleus: the shell model \cite{Cau.98,CNP.01}, 
the quantum Monte Carlo shell model \cite{Uts.99}, the angular 
momentum projected generator coordinate method based on the 
non-relativistic Gogny interaction~\cite{GER.02,ER.03}. 
Although virtually all self-consistent mean-field models, non-relativistic
as well as relativistic, predict a spherical ground state for $^{32}$Mg, 
the GCM calculation with the Gogny force has shown that the ground state 
becomes deformed as a result of the inclusion of rotational energy 
correction. Both the excitation energies $E(2_1^+)$~\cite{Det.84} and 
$E(4_1^+)$~\cite{Klo.93,Gui.02}, as well as the transition probability 
$B(E2,0_1^+\to 2_1^+)$~\cite{Mot.95,Church.05}, 
have been measured for $^{32}$Mg. 
When compared to data from neighboring nuclei, the relatively low 
excitation energy of the first excited state $E(2_1^+) = 885$ keV, 
the large transition probability $B(E2,0_1^+\to 2_1^+)$, and the ratio
$E(4_1^+)/E(2_1^+)=2.6$, indicate that the ground state of $^{32}$Mg is
deformed.   

In Fig.~\ref{figF} we display the mean-field binding energy curves for 
$^{32}$Mg as functions of the quadrupole moment, calculated with the 
PC-F1 relativistic point-coupling effective interaction. 
The constrained mean-field equation has been solved self-consistently 
on a regular mesh ranging from $q=-2.2$b to $q=4.0$b, with the mesh spacing
$\Delta q=0.2$b. The three curves correspond to calculations with
$N_{sh}=8, 10$ and $12$ major oscillator shells. For such a light 
system and for this range of deformations, it appears that 
already $10$ oscillator shells are sufficient to obtain a reasonably 
converged mean-field binding energy curve. In the following calculations
we expand the Dirac spinors in the axially deformed oscillator 
basis with $N_{sh}=10$ major shells. This choice is also supported 
by the results of Ref.~\cite{GER.02}, where correlations beyond the 
mean-field approximation have been studied in the framework of the 
angular momentum projected GCM with the Gogny force. In addition 
to a spherical ground-state, the PC-F1 binding energy curves 
display a prolate deformed shoulder at $q=1.5$b, at an excitation 
energy of $\approx 3.5$ MeV above the ground state. The binding 
energy curve calculated with the Gogny force is similar (see Fig. 6 of 
Ref.~\cite{GER.02}), but the prolate shoulder is somewhat more 
pronounced, and is located only $\approx 1.9$ MeV above 
the spherical ground state. Of course, if the shoulder is too high 
above the spherical ground state, correlations related to the restoration
of rotational symmetry and quadrupole fluctuations might not be strong 
enough to deform the nucleus. The different predictions for the location
of the shoulder can be related to the single-particle levels calculated
with the PC-F1 interaction, displayed in Fig.~\ref{figG}, and with the 
Gogny interaction (Fig. 5 of Ref.~\cite{GER.02}). In these figures the 
eigenvalues of the corresponding mean-field Hamiltonians are plotted 
as functions of the quadrupole deformation. The ratio between the 
neutron spherical gap (7.2 MeV), and the gap at deformation $q=1.5$b 
(2.9 MeV), is $\approx 2.5$ for the PC-F1 interaction, whereas the Gogny 
force gives a much smaller value for this ratio $\approx 1.8$. This leads 
to a more pronounced prolate shoulder at lower excitation energy.  

The essential step in the procedure of angular momentum projection is the
evaluation of the projected norm overlap kernel
\begin{equation}
\mathcal{N}^J(q,q)=\bra{\phi(q)}P^J_{00}\ket{\phi(q)}=
  (2J+1)\frac{1+(-1)^J}{2}\int_0^{\pi/2}{\sin{\beta}d^{J*}_{00}(\beta)n(q;\beta)
     d\beta}\;,
\end{equation} 
where
\begin{equation}
n(q;\beta)=\bra{\phi(q)}e^{-i\beta\hat{J}_y}\ket{\phi(q)}\;.
\label{normbeta}
\end{equation}
In several studies \cite{RG.87,BBH.04,ER.03} it has been shown that the ansatz 
\begin{equation}
n_{app}(q;\beta)=e^{-\frac{1}{2}\langle \hat{J}^2_y\rangle {\rm sin}^2\beta} \;, 
\label{normapp}
\end{equation}
presents an excellent approximation for the
function $n(q;\beta )$, both at small and large deformations. 
The expectation value $\langle \hat{J}^2_y \rangle$ as a function of 
the quadrupole moment is plotted in Fig.~\ref{figH}. This curve is 
in agreement with the one obtained with the Gogny interaction 
(see the right panel in Fig. 5 of Ref.~\cite{ER.03}).  
In Fig.~\ref{figI} we display the function $n(q;\beta)$ for
several values of the quadrupole moment. The solid curves correspond to the 
approximate expression Eq. (\ref{normapp}), whereas dots denote 
values obtained with the exact calculation. The comparison between the 
exact and approximate results provides a very useful test of the numerical 
procedure used in angular momentum projection. 
The projected norm overlap kernels, shown in Fig.~\ref{figJ} for the four
lowest angular momenta, can be compared with those obtained using the Gogny 
effective interaction (see Fig. 7 of Ref.~\cite{GER.02}). We notice that 
the spherical configuration is a pure $0^+$ state 
($\mathcal{N}^{J=0}(0,0) = 1$). The maxima of the projected norm overlap 
kernels for higher angular momenta are correspondingly shifted to larger 
deformations.  

In Fig.~\ref{figK} the energies of the angular momentum projected states 
are analyzed. At this stage we do not consider configuration mixing yet, 
and the projected energy of the $\ket{\phi(q)}$ state reads
\begin{equation}
E^J(q)=\frac{\mathcal{H}^J(q,q)}{\mathcal{N}^J(q,q)}\;.
\label{eproj}
\end{equation}
The angular momentum projected energy curves for 
$J^\pi = 0^+, 2^+, 4^+, 6^+$, and $8^+$ are plotted, together with the
corresponding mean-field binding energy curves, as functions of 
the quadrupole deformation. The curves obtained from solutions in 
axially deformed oscillator bases with $N_{sh}=8, 10$ and
$12$ major shells are almost identical. Since the spherical configuration
is already a pure $0^+$ state, there is no energy gain for $J^\pi = 0^+$ 
at $q=0$b. Notice that the spherical point $q=0$b is not included in plots 
of $E^J(q)$ for $J\ge2$. Namely, for $J \neq 0$ the quantities 
$\mathcal{H}^J(0,0)$ and $\mathcal{N}^J(0,0)$ are so small, that 
their ratio Eq.~(\ref{eproj}) cannot be determined accurately. 
For higher values of the angular momentum 
($J^\pi = 6^+,8^+$ in Fig.~\ref{figK}) several additional configurations 
close to the spherical point are also characterized by very small values 
of the projected norm overlap kernel. These configurations can be safely 
omitted from the projected energy curves, because on the one hand  
the angular momentum projection becomes inaccurate at these points, 
and on the other hand the corresponding angular momentum projected 
states would not play any role in configuration mixing calculations.
 
Is is interesting to compare the projected energy curves with those
obtained using the Gogny effective interaction 
(see Fig. 6 of Ref.~\cite{GER.02}). The principal difference 
is seen already for the $J=0^+$ projected energy. The PC-F1 interaction 
predicts two almost degenerate minima at small oblate and prolate deformations.
The occurrence of degenerate oblate and prolate minima, symmetrical with 
respect to the spherical configuration, is a feature common to all nuclei 
for which the mean-field calculation predicts a spherical ground 
state \cite{ER.03}. As compared to the mean-field energy, the 
prolate deformed shoulder is more pronounced for the $E^{J=0}(q)$ curve, 
and its excitation energy has been lowered from $3.5$ MeV to $1.2$ MeV
by angular momentum projection. On the other hand, at the mean-field level 
the Gogny interaction predicts a more pronounced shoulder, 
only $\approx 1.9$ MeV above the spherical minimum. With angular momentum 
projection the shoulder becomes the absolute minimum of the $J=0$ projected
energy curve. Therefore, the inclusion of the rotational energy correction 
leads to a deformed ground state in $^{32}$Mg, when calculated with the 
Gogny interaction. In addition, the degenerate oblate and prolate minima, 
symmetrical with respect to the $q = 0$, are predicted at slightly higher 
excitation energy. In the present calculation with the PC-F1 interaction, 
the gain in rotational energy is too small to deform the ground state 
of $^{32}$Mg. The rotational energy correction $E_{\rm REC}$, 
i.e. the difference between the mean-field and the $J^\pi=0^+$ 
projected energy curves is plotted in Fig.~\ref{figL}. $E_{\rm REC}$ is 
zero for the spherical intrinsic state, and generally it 
increases rather steeply for small deformations (see also Fig.~\ref{figK}).
Our result for $E_{\rm REC}$ is very similar to the curve obtained 
from the Gogny mean-field potential energy (see Fig. 6 of Ref.~\cite{ER.03}).
This means that the deviation between the $J^\pi=0^+$ projected energy curves
can indeed be attributed to the difference between
the PC-F1 and Gogny interactions on the mean-field level.

\subsection{\label{subIIIc}Angular momentum projection and 
configuration mixing: $^{32}$Mg}

As a final test of our implementation of the GCM for relativistic mean-field
models, we have performed configuration mixing calculations of the angular 
momentum projected intrinsic states for $^{32}$Mg. The solution of the 
Hill-Wheeler matrix equation (\ref{mat_coll}), with the collective 
Hamiltonian Eq.~(\ref{Hcoll}), determines both the ground-state energy, 
and the energies of excited states, for each value of the angular momentum $J$. 
The collective wave functions $g_\alpha^J(q)$, and the weight functions
$f_\alpha^J(q)$, are calculated from the norm overlap eigenfunctions 
Eqs. (\ref{g_u}) and (\ref{f_u}), respectively. As we have shown in 
Fig.~\ref{figJ}, for $J\ge 2$ several points on the energy surfaces
close to the $q = 0$ correspond to configurations with very small values
of the projected norm kernel. Since the numerical evaluation of the 
norm overlap and Hamiltonian kernels is not accurate in such cases, we
have excluded from the configuration mixing calculation all those 
intrinsic configurations for which $\mathcal{N}^J(q,q) < 0.001$ .

The energies and the average quadrupole moments Eq.~(\ref{avquad})
of the two lowest GCM states for each angular momentum are displayed 
in Fig.~\ref{figM}, together with the corresponding 
projected energy curves. The spectrum can be compared 
with the available data, and with the angular momentum 
projected GCM results obtained using the Gogny effective interaction
(see Fig. 7 of Ref.~\cite{ER.03}). Configuration mixing between the 
two essentially degenerate oblate and prolate minima of the $J=0^+$ energy
curve, symmetrical with respect to $q = 0$, results in the almost 
spherical ground state $0_1^+$. When calculated with the Gogny interaction, 
on the other hand, the ground state is prolate deformed. The relatively 
large deformation is a result of a fine balance between the energy 
correction associated with the restoration of rotational symmetry 
(favors larger deformation), and the correlations induced by quadrupole 
fluctuations (mixing between oblate and prolate configurations reduces the 
deformation of the lowest $0^+$ state).

The excitation energies of the $2_1^+$, $4_1^+$ and $6_1^+$ GCM states are 
included in Table~\ref{TabA}, together with the corresponding 
energies obtained with the Gogny force, and the available
experimental excitation energies. Obviously, the PC-F1 interaction 
predicts yrast states at excitation energies that are too high, 
compared with the Gogny interaction, or with the experimental values.
In Table~\ref{TabB} we display the spectroscopic quadrupole moments of the
$2_1^+$, $4_1^+$ and $6_1^+$ GCM states, for the PC-F1 and Gogny
effective interactions. Although comparable in size, 
the quadrupole moments calculated with the PC-F1 interaction are 
systematically smaller. This is,
of course, consistent with the lower excitation energies predicted 
by the Gogny force. Since the ground state is almost spherical, 
the calculated transition probability $B(E2;0_1^+\to 2_1^+) = 
15.5~e^2~fm^4$, is far too small when compared to the experimental 
value ($447(57)~e^2~fm^4$) \cite{Church.05}, or to the value 
obtained with the Gogny interaction ($395~e^2~fm^4$) \cite{ER.03}.

The differences in the spectra predicted by the PC-F1 and Gogny 
interactions originate in the deviation of the corresponding 
mean-field binding energy curves or, more precisely, in the 
different neutron single-particle levels (Nilsson diagrams) 
calculated with the two effective interactions. Because of the  
large spherical gap predicted by the PC-F1 interaction, the magic number 
$N=20$ persists even in such a neutron-rich system,  
and $^{32}$Mg exhibits structure properties typical for other magic nuclei,
e.g. $^{48}$Ca~\cite{ER.03}. This is further illustrated in Fig.~\ref{figN},
where we display the amplitudes of the collective wave functions
$|g_k^J(q)|^2$ for the two lowest GCM states of each angular momentum,
together with the corresponding projected energy curves.
For instance, $|g_1^0(q)|^2$ obviously reflects a configuration mixing of 
the prolate and oblate minima with almost equal weights, resulting in
a ground state with an average quadrupole moment close to zero.
The rotational energy correction for the ground state, i.e. the energy 
gain from angular momentum projection, is $\approx 1$ MeV. 
Configuration mixing provides an additional gain of $0.3$ MeV. 
Both values are in agreement with the corresponding quantities 
calculated for the magic $^{48}$Ca nucleus \cite{ER.03}. 
The amplitudes $|g_1^J(q)|^2$ for $J=2$, $4$ and $6$, are localized 
in the prolate wells of the corresponding projected energy curves and,
therefore, the average quadrupole moments of the states 
$2_1^+$, $4_1^+$ and $6_1^+$ are close to the prolate minima. 
The collective wave functions $g_2^J(q)$, for $J=2$ ,$4$ and $6$, 
correspond to a band based on the $\beta$ vibrational state $0_2^+$. 

\section{\label{secIV}Summary and outlook}		     
The framework of self-consistent relativistic mean-field (RMF) models 
has been very successfully employed in 
analyses of a variety of nuclear structure phenomena, 
not only in nuclei along the valley of $\beta$-stability, 
but also in exotic nuclei with extreme isospin 
values and close to the particle drip lines. Applications 
have reached a level of sophistication and 
accuracy comparable to the non-relativistic 
Hartree-Fock (Bogoliubov) approach based on Skyrme or 
Gogny effective interactions. Although mean-field and pairing 
correlations are treated very carefully in modern RMF models, 
additional correlations, related to the restoration of broken 
symmetries and to fluctuations, have either been neglected or 
taken into account in an implicit way. In this work we have 
introduced a model in which restoration of rotational 
symmetry and fluctuations of the quadrupole deformation 
are explicitly included in the relativistic framework.

In the specific model which has been developed in this work, 
the generator coordinate method (GCM) is employed to perform
configuration mixing calculations of angular momentum projected
wave functions, calculated in a relativistic point-coupling model.
The geometry has been restricted to axially symmetric shapes, 
and the mass quadrupole moment is used as the generating coordinate.
The intrinsic wave functions are generated from the solutions of 
the constrained relativistic mean-field + BCS equations in an 
axially deformed oscillator basis. 

In order to test our implementation of the GCM and angular 
momentum projection, a number of illustrative calculations 
have been carried out for the nuclei $^{194}$Hg and $^{32}$Mg.
The PC-F1 parameter set \cite{BMM.02} has been used for the 
effective point-coupling Lagrangian, and the effective 
interaction in the particle-particle channel has been 
approximated by a density-independent $\delta$-force. 
The test of the generator coordinate method has been performed 
in a study of quadrupole dynamics in the nucleus $^{194}$Hg, 
and the results have been compared with the classical analysis of 
the GCM in the investigation of shape isomerism in $^{194}$Hg by Bonche 
{\it et al.} \cite{Bon.90}, based on the non-relativistic 
constrained Hartree-Fock plus BCS model with the SIII Skyrme 
effective interaction. Angular momentum projection and, finally, 
configuration mixing of angular momentum projected states, have 
been tested in the example of the neutron-rich nucleus $^{32}$Mg,
in comparison with results obtained with the angular 
momentum projected generator coordinate method based on the 
non-relativistic HFB with the Gogny interaction~\cite{GER.02,ER.03}.
The tests have been very successful, and the results obtained 
for the binding energy curves, projected energy curves, rotational 
energy corrections, ground and low-lying excited states, and 
collective wave functions for $^{194}$Hg and $^{32}$Mg, are generally 
in very good agreement with the predictions of GCM calculations based on 
non-relativistic Skyrme and Gogny interactions, respectively.

The choice of the PC-F1 relativistic effective interaction, however, 
does not lead to a deformed solution for the ground state of 
$^{32}$Mg, even after the inclusion of the rotational energy 
correction. This result is in contrast with available data, and 
with the configuration mixing calculation of angular momentum 
projected configurations based on the Gogny interaction. The 
different predictions for the ground state of $^{32}$Mg can be 
related to the corresponding mean-field binding energy curves and,
more specifically, to the different results for the size of the spherical 
$N=20$ neutron gap, obtained with the Gogny and PC-F1 interactions. 
Even though the spherical ground state of $^{32}$Mg, predicted by 
the PC-F1 effective interaction, is not crucial in the context of 
the present analysis, it points to an important problem. Namely, 
the choice of effective interactions to be used in self-consistent
calculations that go beyond the mean-field approximation and 
explictly include correlations, such as those considered in the 
present work. Virtually all global effective interactions have 
been adjusted to data, e.g. masses and radii, which already include
correlations. On the other hand, those correlations that we wish 
to treat explicitly, should not be included in the effective 
interaction in an implicit way. The solution is to adjust global 
effective interactions to pseudodata, obtained by subtracting 
correlation effects from experimental masses and radii. Approximate
methods for the calculation of correlations have recently been 
developed \cite{BBH.04}, that will enable a systematic evaluation 
of correlation energies for the nuclear mass table.      

Before proceeding with realistic applications of the model introduced 
in this work, our first task is to adjust a new global effective 
point-coupling interaction which will not implicitly contain rotational 
energy corrections and quadrupole fluctuation correlations. Further
developments will include the treatment of pairing fluctuations by 
particle number projection, the use of different generating coordinates 
for the neutron and proton density distributions, the description 
of non-axial shapes, and the extension to odd nuclei.


\bigskip 
\noindent
\bigskip \bigskip

\bigskip \bigskip
\leftline{\bf ACKNOWLEDGMENTS}
This work has been supported in part by the Bundesministerium
f\"ur Bildung und Forschung - project 06 MT 193, by 
the Alexander von Humboldt Stiftung, 
and by the Croatian Ministry of Science - project 0119250.

\newpage
\begin{figure}
\caption{The binding energy curves for $^{194}$Hg,
as functions of the mass quadrupole moment, calculated by expanding the 
Dirac spinors in $10, 12, 14, 16$ and $18$ oscillator shells.}
\label{figA}
\end{figure}

\begin{figure}
\caption{The eigenvalues of the norm overlap kernel, calculated 
using four different values for the mesh spacing: 
$\Delta q=2, 4, 6$ and $8$ b (left panel).
The energies of the twelve lowest GCM states plotted as functions of 
the dimension of the GCM basis (right panel).}
\label{figB}
\end{figure}

\begin{figure}
\caption{The energies of fifteen lowest GCM states in $^{194}$Hg, plotted 
as functions of the average quadrupole moment, together with the corresponding
mean-field binding energy curves. The four panels correspond to calculations 
in oscillator bases with $10, 12, 14$ and $16$ major oscillator shells. }
\label{figC}
\end{figure}

\begin{figure}
\caption{GCM collective wave functions $g_k(q)$ Eq.~(\protect\ref{coll_wf}) 
for the lowest fourteen states in $^{194}$Hg. The vertical dashed line 
denotes the position of the barrier separating the main and the 
superdeformed potential wells.}
\label{figD}
\end{figure}

\begin{figure}
\caption{The mean-field binding energy curves for $^{194}$Hg, 
together with the energies and average quadrupole moments of fifteen lowest 
GCM states. Calculations have been performed using four values of the mesh 
spacing: $\Delta q=2, 4, 6$ and $8$ b. Zero energy is placed at 
the position of the minimum of the binding energy curve. }
\label{figE}
\end{figure}

\begin{figure}
\caption{The binding energy curves for $^{32}$Mg, calculated from the 
constrained solutions of the self-consistent relativistic mean-field
equations in axially deformed oscillator bases with 8, 10 and 12
major shells.}
\label{figF}
\end{figure}

\begin{figure}
\caption{The neutron (left panel) and proton (right panel) single-particle
levels for $^{32}$Mg, as functions of the mass quadrupole moment. The thick 
dashed curve denotes the position of the Fermi energy.}
\label{figG}
\end{figure}

\begin{figure}
\caption{The expectation value $\langle \hat{J}_y^2 \rangle$ for $^{32}$Mg, 
as a function of the mass quadrupole moment.}
\label{figH}
\end{figure}

\begin{figure}
\caption{A comparison between the exact values of the function $n(q;\beta)$
Eq.~(\protect\ref{normbeta}) (dots) and the approximate expression 
Eq.~(\protect\ref{normapp}) (curves), for several 
values of the mass quadrupole moment.}
\label{figI}
\end{figure}

\begin{figure}
\caption{Projected norm overlap kernel $\mathcal{N}^J(q,q)$ as a function of 
the mass quadrupole moment for $^{32}$Mg.}
\label{figJ}
\end{figure}

\begin{figure}
\caption{Angular momentum projected ($J^\pi=0^+ ,2^+ ,4^+ ,6^+$ and $8^+$)
potential energy curves for $^{32}$Mg, as functions of the mass quadrupole
moment. The mean-field energies are also included (thick dotted curves). 
The three panels correspond to solutions in axially deformed oscillator bases
with 8, 10 and 12 major shells.}
\label{figK}
\end{figure}

\begin{figure}
\caption{Rotational energy correction as a function of the mass 
quadrupole moment for $^{32}$Mg.}
\label{figL}
\end{figure}

\begin{figure}
\caption{The energies and the average quadrupole moments
of the two lowest GCM states for each angular momentum in 
$^{32}$Mg, together with the corresponding 
projected energy curves.}
\label{figM}
\end{figure}

\begin{figure}
\caption{Squares of the collective wave functions 
$|g_k^J(q)|^2$ of the two lowest GCM states for 
each value of the angular momentum in $^{32}$Mg,
together with the corresponding 
projected energy curves.}
\label{figN}

\end{figure}
\newpage

\begin {table}[]
\begin {center}
\caption {The excitation energies (in MeV) 
of the $2_1^+$, $4_1^+$ and $6_1^+$ GCM states
for $^{32}$Mg.}
\bigskip
\begin {tabular}{cccc}
\hline
\hline
     &{ E (PC-F1)}  & { E (Gogny)} & {E (exp.)}   \\ 
\hline
$2_1^+$    & $2.04$   &  $1.4$ &  $0.885$       \\
\hline
$4_1^+$    &  $4.42$  &  $3.6$ &  $1.437$       \\
\hline
$6_1^+$    & $7.41$   &  $5.5$ &       \\
\hline
\hline
\end{tabular}
\label{TabA}
\end{center}
\end{table}

\begin {table}[]
\begin {center}
\caption {The spectroscopic quadrupole moments 
(in $e~fm^2$) of the $2_1^+$, $4_1^+$ and $6_1^+$
GCM states for $^{32}$Mg.}
\bigskip
\begin {tabular}{ccc}
\hline
\hline
     &{ $Q_2^{spec} $ (PC-F1)}  & { $Q_2^{spec} $ (Gogny)}   \\ 
\hline
$2_1^+$    & $-17.51$   &  $-19.15$      \\
\hline
$4_1^+$    &  $-19.28$  &  $-26.31$    \\
\hline
$6_1^+$    & $-21.19$   &  $-30.09$      \\
\hline
\hline
\end{tabular}
\label{TabB}
\end{center}
\end{table}
\end{document}